\documentstyle[12pt,a41,epsfig,wrapfig,rotating]{article}

\begin{document}
\begin{flushleft}
DESY 98-183\\
hep-ph/9811361\\
November 1998
\end{flushleft}
\vspace*{4cm}

\begin{center}
{\Large \bf  $\Upsilon$ Polarization at HERA-$B$}

\vspace*{2cm}
A.~Kharchilava$^a$, T.~Lohse$^b$, A.~Somov$^b$, A.~Tkabladze$^c$\footnote{Alexander von Humboldt Fellow}

\vspace*{1cm}

{\small \it $^a$ Institute of Physics, Georgian Academy of Sciences, Tbilisi} \\
{\small \it $^b$ Humboldt-University, Berlin, Germany } \\
{\small \it $^c$ DESY Zeuthen, D-15738 Zeuthen, Germany} \\  

\vspace*{2.5cm}

\end{center}

\begin{abstract}
{\small
The production of $\Upsilon$ mesons in  fixed target $pN$ collisions 
is considered. It is shown that Non-Relativistic QCD
predicts $\Upsilon$ states to be produced with sizeable transverse 
polarization. The possibility of a measurement of the $\Upsilon$ polarization
at the HERA-$B$ experiment is discussed.
}
\end{abstract}
\vspace*{0.5cm}

\noindent
\hspace*{1cm}{\it PACS}: 13.20.Gd, 13.88.+e, 13.60.Le, 12.38.Qk

\newpage

\vspace{0.3cm}

\section{Introduction}
\setcounter{equation}{0}
\vspace{-3mm}

\noindent
The Factorization Approach (FA) based on the Non-Relativistic QCD (NRQCD) 
represents a reliable framework to study heavy quarkonium production
and decay processes \cite{NRQCD}. According to the FA the
inclusive production cross section for a quarkonium state
 $H$ in the process 
\begin{eqnarray}
 A+B\to H+X
\end{eqnarray}
can be factorized as
\begin{eqnarray}
\sigma(A+B\to H) = 
\sum_{n}{\frac{F_n}{m_Q^{d_n-4}}\langle0|{\cal O}^H_n|0\rangle},
\end{eqnarray}
where the short-distance coefficients, $F_n$, are associated with the
production of a  heavy quark pair in the color and angular momentum state 
$[n]$. This part of the  cross section involves only momenta at least 
of the order of the heavy quark mass, $m_Q$,  and  can be calculated
 perturbatively. Two distinct scales are introduced: the heavy 
quark-antiquark pair production process occurs at  small distances, 
$~1/(m_Q)$, and is  factorized from the hadronization phase which 
takes place at large  distances, $1/(m_Q v^2)$.
Here $v$ is the average velocity
of heavy constituents in the quarkonium, with  $v^2\simeq0.3$ for charmonium 
and $v^2\simeq0.1$ for bottomonium systems. 
The vacuum matrix elements of NRQCD operators,
 $\langle0|{\cal O}^H_n|0\rangle$, describe the evolution of the 
quark-antiquark state $[n]$ into the final hadronic state $H$ \cite{NRQCD}.
 These long distance matrix elements cannot
be calculated perturbatively, but their relative importance
 in powers of velocity $v$ can be estimated  using the NRQCD velocity
 scaling rules \cite{LMNMH}.
The important feature of this  formalism is that the cross section of heavy quarkonium production can be organized as  double expansion in powers of
 $v$ and $\alpha_s(m_Q)$. In  higher order of $v$ the FA implies 
 that the quark-antiquark color octet intermediate  states are allowed
to contribute to heavy quarkonium  production and decay
processes.

 Unlike the color singlet long distance matrix elements, each connected
with the subsequent  non-relativistic wave function at the origin, the 
color octet long distance matrix elements are unknown and have to be 
extracted from the experimental data.
The NRQCD factorization approach implies universality, i.e. the values
of long distance matrix elements extracted from  different
experimental data sets must be the same. 
However, due to the presently rather large theoretical 
uncertainties \cite{Trig,BK} and the  unknown size of  higher twist 
process contributions \cite{VHBT} the existing  experimental data do
not allow yet to check the FA universality.

The data on direct $J/\psi$ and $\psi'$ production at large transverse momenta
at the Tevatron indicates that the color octet contribution
is dominating. The  $S$ state charmonia are produced through the
 gluon fragmentation into  the $^3S_1^{(8)}$ octet state \cite{BF,CL}.
Recent investigations have shown  that  the contribution of color 
octet states  to the charmonium and bottomonium production cross sections 
is very important at fixed 
target energies, $\sqrt{s}\simeq30-60$ GeV,
 and reduces existing discrepancies between experimental data and 
predictions of the Color Singlet Model (CSM) \cite{FixTag,BR}.
However,  some experimental data contradict 
 the Color Octet Model (COM) predictions.
 In particular, theoretical
predictions disagree  with measurements of the polarization of
   $J/\psi$ and $\psi'$ particles produced at
fixed target energies \cite{BR} and  the COM predicts a too
low relative yield of the $\chi_{c1}$ state compared to the $\chi_{c2}$
 \cite{BR}.
One  possible solution of   these discrepancies was proposed
by Brodsky et al. \cite{VHBT}  suggesting that higher twist 
processes, when more than one parton from projectile or target participates
in the reaction, might give a significant 
contribution to low $p_T$ production of $J/\psi$ and $\chi_{c1}$ states.
Problems exist also in charmonium photoproduction at HERA.
The color octet contribution underestimates the inelastic $J/\psi$ 
photoproduction cross section at large values of $z$
($z=E_{J/\psi}/E_{\gamma}$ in the laboratory frame) \cite{CK}.

Spin effects in heavy quarkonia production 
are expected to provide  tests for the
different mechanisms of heavy quarkonium production \cite{VHBT,BK,TT,NT}.
Predictions for the polarization of direct $\psi$'s produced at large $p_T$ 
at the Tevatron are free from theoretical uncertainties connected with higher 
twist effects and corresponding  measurements will  provide an
excellent possibility   to test the NRQCD factorization approach.
The observation of opposite  sign  double spin asymmetries in
the production of different charmonium states 
can also be used to discriminate 
the NRQCD FA from  the Color Evaporation Model (CEM) which predicts 
the same spin asymmetries for all charmonium states \cite{TT,NT}.
The CEM also assumes a factorization between the production of a heavy
quark pair and its
hadronization phase. But unlike the NRQCD factorization approach
the CEM postulates that multiple soft gluon exchange in the hadronization phase
destroys the initial polarization of the heavy quark pair and the heavy
quarkonium is produced unpolarized \cite{CEM}.

At the same time it is not excluded that the mass of the charm quark 
is not large enough to apply the NRQCD factorization approach to
charmonium production and  decay processes. 
Due to the rather  large value of $v^2$, 
about $0.3$ for the charmonium system, the Fock states at higher order
of $v^2$ may give the essential contribution and can then not be  neglected.
When fitting the values of the long distance parameters the
heavy quark spin symmetries are  used to reduce the number of independent
parameters. These relations are valid up to $v^2$ and may  get 
large corrections for charmonium production.
In contrast, the NRQCD FA predictions for the bottomonium system are more 
reliable, since the expansion parameter $v^2$ is much smaller (around $0.1$),
than for the charmonium system.
Higher twist processes are also  expected to be suppressed as
 $\Lambda/m_b\simeq0.1$ (compare with $\Lambda/m_c\simeq0.3$).
Also the QCD coupling constant is smaller  for bottomonium system.
Therefore, the  characteristics of $\Upsilon$ meson
production are more appropriate for a correct test of 
the  NRQCD factorization approach. 

In this article we consider the polarization of $\Upsilon$ mesons  
produced at fixed target energies. It will be shown that in the NRQCD 
factorization approach $\Upsilon$ mesons are produced transversely
polarized, whereas the
CEM predicts unpolarized bottomonium production. 
We present also numerical estimates of the projected statistical errors 
for the  measurement of $\Upsilon$ polarization at the 
HERA-$B$ experiment at DESY.

In the next section the bottomonium production in the various
subprocesses is  discussed. In section 3 the $\Upsilon$ polarization is 
calculated and numerical estimates for the expected $\Upsilon$ signal 
to background ratio as well as for the errors of the polarization
measurement  are considered in the section 4.

\vspace{0.3cm}

\noindent
\section{ $\Upsilon$ Production Subprocesses and Matrix Elements.}
In leading order in $\alpha_s$ the different $S$- and $P$-wave 
quark-antiquark states can be produced in the following $2\to2$ and $2\to3$
subprocesses:

$\bullet$ gluon-gluon fusion
\begin{eqnarray}
& gg\rightarrow\;^1\!S^{(1,8)}_0 & \hspace{0.5cm} 
                         gg\rightarrow\;^3\!S^{(1,8)}_1+g \nonumber\\
& gg\rightarrow\;^3\!P^{(1,8)}_{0,2} & \hspace{0.5cm}  
                         gg\rightarrow\;^3\!P^{(1,8)}_1+g
\end{eqnarray}

$\bullet$ gluon-quark scattering
\begin{eqnarray}
&gq\rightarrow\;^3\!S^{(8)}_1+q \hspace{0.5cm} & 
gq\rightarrow\;^3P^{(1,8)}_1+q
\end{eqnarray}

$\bullet$ quark-antiquark annihilation
\begin{eqnarray}
&&q\bar q\rightarrow\;^3\!S^{(8)}_1
\end{eqnarray}
where the superscripts (1,8) denote the color singlet and color octet
states, respectively.

The total cross section of $\Upsilon$ production is given by the sum
of direct production cross section and the cross section of $\chi_{bJ}$ 
states decaying through $\Upsilon$ mesons:
\begin{eqnarray}
\sigma_{\Upsilon} = \sigma(\Upsilon)_{dir}
+\sum_{J=0,1,2}{Br(\chi_{bJ}\to\Upsilon X)\sigma(\chi_{bJ})}+
\sum_{n=2,3}{Br(\Upsilon(n)\to\Upsilon X)\sigma(\Upsilon(n))}.
\end{eqnarray}
The production of each quarkonium state receives  contributions from both  
color octet and color singlet states.
The relative velocity for the bottomonium system is small, $v^2\simeq0.1$,
and hence the production cross sections can be calculated with only the
leading order color octet contribution taken into account.

In table I the color singlet and color octet long distance matrix elements 
for the production of $\chi_{bJ}$ states are presented.
The values of matrix elements are taken from \cite{BR}. The color singlet
matrix elements are computed from the wave functions of 
the Buchm\"uller-Tye potential tabulated in \cite{EQ}.
The matrix elements $\langle{\cal O}_8^H(^3S_1)\rangle$ are fitted from the 
Tevatron data \cite{CL}. For the $3P$ bottomonium state the value of this 
matrix element is obtained by extrapolation from $1P$ and $2P$
states \cite{BR}. 

\begin{center}
\begin{tabular}{|c|c|c|c|}
\hline \hline
 Matrix Elements & $\chi_{b0}(1P)$ & $\chi_{b0}(2P)$ & $\chi_{b0}(3P)$\\
\hline
$\langle{\cal O}_1^{H}(^3P_0)\rangle/m_b^2$ & $8.5\cdot10^{-2}$ & 
$9.9\cdot10^{-2}$ & $0.11$ \\
$\langle{\cal O}_8^H(^3S_1)\rangle$  & $0.42$ & $0.32$ & $0.25$ \\
\hline
\end{tabular}
\end{center}

\noindent
{\bf Table I.} Matrix elements for the $P$-wave bottomonia production in 
units GeV$^3$.
\vspace{0.3cm}
 
The characteristics of $\Upsilon$ production at fixed target energies 
($\sqrt{s}\simeq40$ GeV) differ from charmonium production, as it was mentioned
in \cite{BR}. The $\Upsilon(nS)$ states are mainly produced 
in the quark-antiquark annihilation subprocess 
through decays of $P$ wave states of bottomonia.
Unlike $J/\psi$ production, the color octet contribution to
the  $\Upsilon$ production in the gluon-gluon fusion subprocesses is less
important. 
In leading order of perturbative QCD (pQCD) only  $^1S_0^{(8)}$
and $^3P_J^{(8)}$ color octet states contribute  to $\Upsilon$ production.
They are  suppressed as $v^4$ and $v^2$ 
compared to $\Upsilon$
production through color singlet $^3S_1$ and $^3P_J$ states,
respectively. Hence, due to the small value of $v^2\simeq0.1$, it turns 
out that
in the gluon-gluon fusion subprocesses the main contribution to $\Upsilon$
production comes from the color singlet states.
As can be seen from (3), the $P$-wave bottomonium states are  produced in 
 lowest order in pQCD expansion, ${\cal O}(\alpha_s^2)$.
The subprocesses for direct production of $S$ states ($\Upsilon(nS))$
are suppressed by the factor $\alpha_s/\pi$ compared to those for $\chi_{b0}$ 
and $\chi_{b2}$  production.
Consequently, the indirect $\Upsilon$ meson fraction from the decay of
$\chi_{bJ}$ states is expected to be large in the gluon-gluon fusion 
subprocesses.

The main contribution to $P$-wave bottomonia production comes from
the  quark-antiquark annihilation subprocess.
First of all,
the  leading color octet and color singlet
contributions to $P$-wave quarkonium production scale equally in $v^2$, 
$O(v^5)$,  the subleading corrections being only of the order $O(v^9)$.
In leading order of   $v^2$,  one color octet state contributes
to the production of $\chi_{bJ}$ states, namely $^3S_1^{(8)}$.
Moreover, in leading order of pQCD only this state is produced 
in the quark-antiquark annihilation subprocesses.
The  values of the matrix elements 
$\langle{\cal O}^{\chi_{bJ}}_8(^3S_1)\rangle$ are larger as compared to the
color singlet matrix elements for corresponding $P$ states,
connected to the quarkonium wave functions derivatives at the origin 
(see table I) \cite{CL}. 
Furthermore, at fixed target energies the $q\bar q$ luminosity effectively 
increases compared to the $gg$ luminosity due to the 
large mass of the $(b\bar b)$ system.

The $\Upsilon$ production cross sections   in gluon-gluon
fusion and quark-antiquark annihilation subprocesses are presented in 
table II. The  $b$-quark mass of $m_b=4.9$ GeV is chosen
 as in ref. \cite{CL} in order to extract the values of color octet long 
distance parameters from the Tevatron data for the
$\Upsilon$ production. The cross sections are calculated using the GRV LO
parton distribution functions \cite{GRV} evaluated at the factorization
scale $Q^2 = 4 m_b^2$.
The long distance color octet parameters for direct $\Upsilon(nS)$ 
production are taken from \cite{BR}. 
For the decays of $\chi_{bJ}(3n)$ states\footnote{As was suggested 
in \cite{BR}, the  large yield of the  $\Upsilon(3S)$ state at the 
E772 experiment  at FNAL \cite{exp1} could  be explained by assuming 
the unobserved $\chi_{bJ}(3P)$ states to lie below the open beauty threshold.}
to $\Upsilon(3S)$ the same branching ratios are assumed as for 
the corresponding $n=2$ states \cite{BR}.

\begin{center}
\begin{tabular}{|c|c|c|c|}
\hline \hline
subprocess  & \multicolumn{2}{c}{gluon-gluon fusion }\vline&
quark-antiquark annihilation\\
\cline{2-3} 
 & through $\Upsilon(nS)$ & through $\chi_{bJ}$ & through $\chi_{bJ}$  \\
\hline
 $\sigma(\Upsilon)$  & $0.013~nb$  & $0.016~nb$
 & $0.24~nb$ \\
\hline 
\end{tabular}
\end{center}

\noindent
{\bf Table II.} $\Upsilon(1S)$ production cross sections in different 
subprocesses.
\vspace{0.3cm}

From table II one finds that the direct $\Upsilon$ production cross 
section for the adopted values of the long distance parameters is  
more than one order of magnitude  smaller than the total cross section.
Even in the color singlet model the $\Upsilon$ mesons are mainly produced
through decays of $\chi_{bJ}$ states. On the other hand, 
the color octet contribution seems to be dominant in $\chi_{bJ}$ 
production. The main contribution comes from the $^3S_1^{(8)}$ state,
produced in quark-antiquark annihilation.

As was already shown in \cite{BR}, the color octet contribution reduces 
the large discrepancy between the CSM prediction for the total $\Upsilon$ 
production cross section and experimental data \cite{exp1,exp2,exp3}. 
Nevertheless, there remain large uncertainties due to contradicting experimental results.
The cross sections obtained by integration of $x_F$ distributions for
$\Upsilon(1S)$ production presented in \cite{exp1} and \cite{exp2} are 
three and four times smaller, respectively,  than the central value quoted in 
\cite{exp3}, $270~pb/nucleon$.  
In addition, the theoretical value of the cross section strongly  
depends  on the assumed  mass of the $b$-quark. It is therefore impossible
to extract the color octet matrix elements with reasonable accuracy 
from the fixed target bottomonia production data.

\section{$\Upsilon$ Polarization}

\noindent
As it was already mentioned in \cite{BR}, the measurement of the cross
section for direct and 
indirect production  of $\Upsilon'$s  would provide crucial information
about the color octet mechanism, e.g., new constraints on the long distance
color octet parameters would emerge.
Such a measurement requires the  reconstruction of $\chi_{bJ}$ states in  
the $\Upsilon+\gamma$ decay mode which is not a trivial task  at fixed target
experiments due to the small transverse momentum of the emitted photon.

Another possibility to check the NRQCD factorization approach is to
measure the  $\Upsilon$ polarization.
In $H \to \ell^+\ell^-$ decays the polarization of $S$-state 
quarkonium  is determined by the polar-angle distribution of its decay
leptons with respect to the beam direction in the meson rest frame.
Integrating  over the azimuthal angle the distribution has the form
\begin{eqnarray}
\frac{d\sigma}{d\cos{\theta}}\propto1+\alpha\cos^2{\theta},
\end{eqnarray}
where $\theta$ is the angle between the positively charged lepton, 
$\ell^+ \ (\ell=$ e, $\mu)$, and the beam axis in the quarkonium rest
frame. The parameter $\alpha$ in the angular distribution can be related
to $\xi$, the fraction of longitudinally polarized  $\Upsilon$ mesons:
\begin{eqnarray}
\alpha = \frac{1-3\xi}{\ 1+\xi}=\cases{\;\;\;1\ {\rm for}\ \xi =0 \cr
-1\ {\rm for}\ \xi = 1\cr}
\end{eqnarray}

The calculation of  $J/\psi$ and $\psi'$ polarization at fixed target 
energies and Tevatron collider energies was performed in \cite{BK,BR,BRp,TV}.
The most general method to calculate the cross sections for
heavy quarkonium production with definite polarization within the NRQCD 
factorization approach was proposed by Braaten and Chen \cite{BC}.
Tang and V\"anttinen used the covariant projection method to calculate cross 
sections for polarized $J/\psi$ and $\psi'$ production \cite{TV}.
The $\chi_{bJ}(nP)$ states are produced in quark-antiquark annihilation
through only one color octet state, $^3S_1^{(8)}$.
For this particular case   both methods  give the same result. 
We used the formulae derived in \cite{TV} to obtain  the 
polarization of $\Upsilon(nS)$ states produced in cascade,
$q\bar q\to\ ^3\!S_1^{(8)}\to\chi_{bJ}\to\Upsilon(nS)+\gamma$:
\begin{eqnarray}
&&\sigma(q\bar q\to b\bar b (^3S_1^{(8)})\to\chi_{b1}+g\to\Upsilon+\gamma+g)
\nonumber\\
&&=\frac{16\pi^3\alpha_s^2}{27 M^5}\delta(1-M^2/\hat s)
 Br(\chi_{1b}\to\Upsilon+\gamma) 
\langle{\cal O}_8^{\chi_{b1}}(^3S_1)\rangle\frac{3-\delta_{\lambda0}}{8},
\end{eqnarray}
\begin{eqnarray}
&&\sigma(q\bar q\to b\bar b (^3S_1^{(8)})\to\chi_{b2}+g\to\Upsilon+\gamma+g)
\nonumber\\
&&=\frac{16\pi^3\alpha_s^2}{27 M^5}\delta(1-M^2/\hat s)
 Br(\chi_{b2}\to\Upsilon+\gamma) 
\langle{\cal O}_8^{\chi_{b2}}(^3S_1)\rangle\frac{47-21\delta_{\lambda0}}{120}.
\end{eqnarray}
The scalar $\chi_{b0}$ state yields unpolarized bottomonium $S$-wave  states.
It follows from (9) and (10) that  $\chi_{b1}$ and $\chi_{b2}$ 
intermediate states lead to  values of $\alpha=0.2$ and $\alpha=0.29$,
respectively. 
Taking into account all transitions from $\chi_{bJ}(1P)$ and $\chi_{bJ}(2P)$
states to $\Upsilon(1S)$ we obtain
for the polarization parameter in quark-antiquark annihilation subprocess 
\begin{eqnarray}
\alpha\simeq0.24.
\end{eqnarray}
 This result remains practically unchanged if we add 
contribution from $\chi_{bJ}(3P)$ states with the same branching ratios as
for the corresponding $2P$ states.
It is worth mentioning that the value $\alpha\simeq0.24$ represents only
a lower  theoretical
limit for the polarization. This value is calculated taking into account only
the quark-antiquark annihilation subprocesses. 
In  gluon-gluon fusion subprocesses the  polarization of $\Upsilon$ mesons is 
larger due to the dominant contribution from $\chi_{b2}$ decays
which yield pure transverse polarization \cite{VHBT}.
For the values of octet long distance parameters  (table I) the size of  
the polarization is $\alpha\simeq0.3$ and hence also exceeds $0.24$.
However, large uncertainties in color octet long distance parameters for 
$P$-wave  bottomonia does not allow 
to compute the relative importance of the various subprocesses, so that the
lower bound of $\alpha=0.24$ remains as the only firm prediction.

In contrast to charmonium 
production the higher twist effects for the bottomonium system are expected 
to be small. In particular, the higher twist effect suggested in 
\cite{VHBT}, when more than one parton from the projectile or target
is involved  in the heavy quarkonium states production, is expected
to be negligible for  $\Upsilon$  production. 
Two partons should be within a transverse distance of $O(1/m_Q)$ in order
to interact with the other parton and to produce a heavy quark-antiquark bound 
state.
Consequently, such a higher twist  process is  suppressed by a factor 
of $O(\Lambda^2_{QCD}/m_Q^2)$ \cite{VHBT}. 
To explain the discrepancies between the CSM predictions and
measured  relative production rate 
 $\chi_{c1}/\chi_{c2}$  it is suggested that 
the above suppression can be compensated  by a kinematical
enhancement \cite{VHBT}. Thus the  $\chi_{c1}$ production cross section in 
the higher twist process is expected to be at the same level as the $\chi_{c2}$
production cross section from gluon-gluon fusion subprocesses.  
In bottomonium production such a mechanism
 is suppressed by the mass of the bottom quark
 $O(\Lambda_{QCD}^2/m_b^2)$ , i.e. the corresponding cross section
should be  one order of magnitude  smaller than in 
the charmonium case. The effect of the kinematical enhancement is also smaller
due to the larger mass of the bottomonium system. Moreover,
as can be seen from table II, the gluon-gluon fusion subprocesses
give small contributions to $\Upsilon$ production at fixed target 
energies compared to $J/\psi$ production.
Therefore, the  measurement of $\Upsilon$ polarization at fixed target
energies will allow to distinguish between the NRQCD approach and the CEM, 
which predicts unpolarized production of all bottomonium states. 


\section{Expectations for HERA-$B$} 

\noindent
HERA-$B$ is an experiment presently set up at DESY which uses the
HERA $920$ GeV/c proton beam incident on various nuclear targets \cite{HERAB}.
Being optimized for the study of the various aspects of $B$-physics,
i.g. the measurement of the CP asymmetry in the $B^0_d \to J/\psi K^0_S$
decays,
the experiment is well suited to perform accurate $\Upsilon$ polarization 
measurements due to the following reasons:  

$\bullet$ large acceptance of the apparatus (polar angle coverage in the
laboratory frame from \linebreak 10 mrad up to 250 mrad),

$\bullet$ good momentum/mass resolution which allows to separate the
$\Upsilon(nS)$ states,

$\bullet$ high statistics due to high interaction rates (40 MHz).

In the current analysis the acceptance for $\Upsilon$ production and the
mass resolution in
the $\Upsilon$ mass region are determined using a detailed
HERA-$B$ detector simulation \cite{hbgeant}. The $\Upsilon$ mesons are
generated
with the PYTHIA 5.7 event generator \cite{pythia} and then processed
through
the simulation of the full detector. The cross sections
for $\Upsilon$ production are taken
from the measurements of the E605 experiment ($\sqrt{s}\simeq38.8$ GeV)
\cite{exp2}:
\begin{eqnarray}
&&Br\frac{d\sigma}{dy}|_{y=0}(\Upsilon(1S)+\Upsilon(2S)+
      \Upsilon(3S))=2.31\; pb/nucleon;\nonumber\\
&&
\frac{Br\frac{d\sigma}{dy}(\Upsilon(2S))}   {Br\frac{d\sigma}
{dy}(\Upsilon(1S))}=0.31; 
\hspace{0.3cm}
\hspace{0.3cm} \frac{Br\frac{d\sigma}{dy}(\Upsilon(3S))}
{Br\frac{d\sigma}{dy}(\Upsilon(1S))}=0.09.
\end{eqnarray}
As it was noted above (in section 2), this experiment gives
the lowest lying value for the $\Upsilon$ production cross section
, about $130~ pb/nucleon$ for $\sigma(\Upsilon(1S)+\Upsilon(2S)+
\Upsilon(3S)).$ 
\begin{figure}[ht]
\begin{minipage}[c]{7.5cm}
\epsfig{file=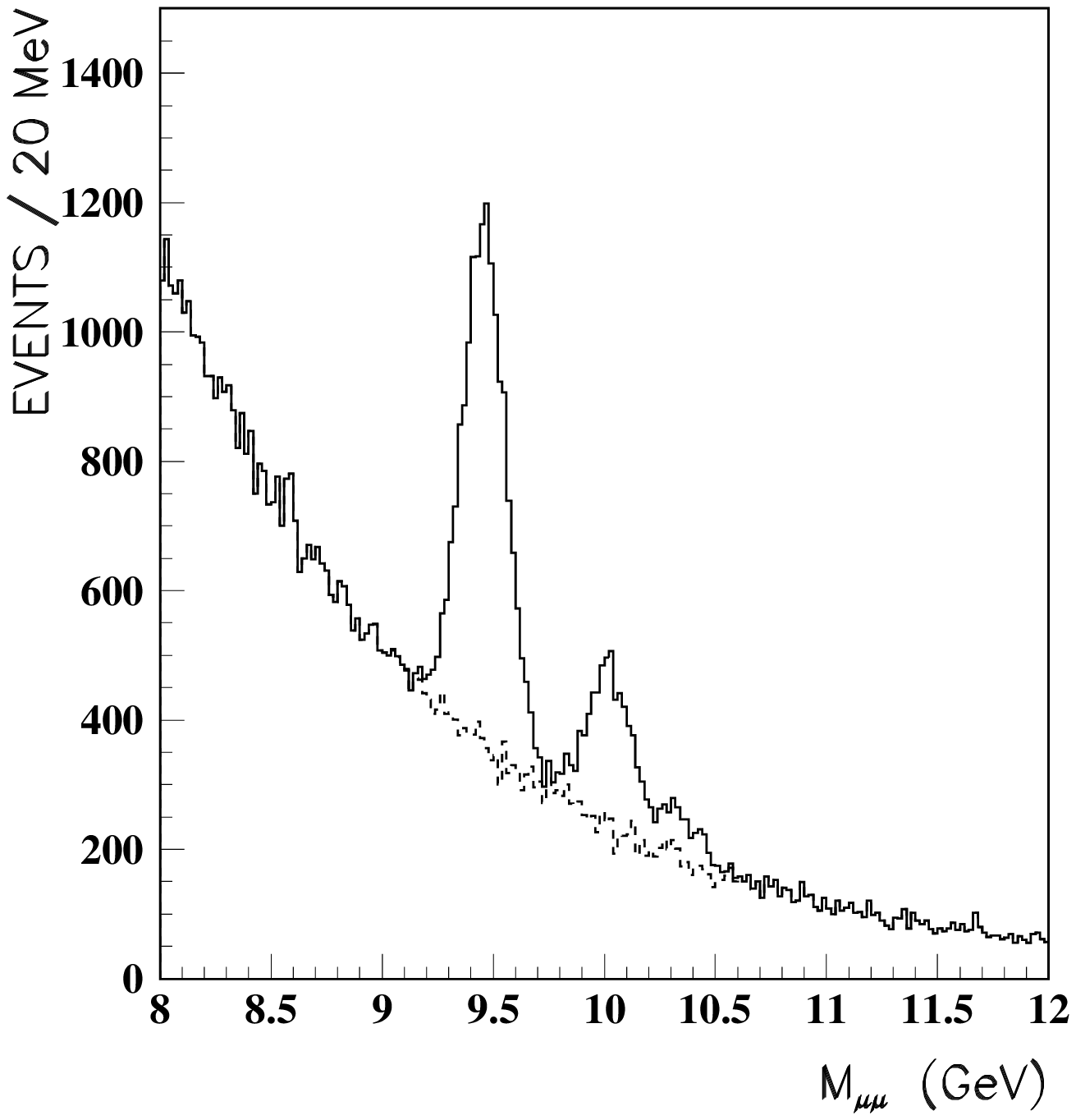,height=7.5cm,width=7.5cm}
\caption {$\Upsilon \to \mu^+\mu^-$ mass distribution in one year running
of HERA-$B$}
\end{minipage}
\hspace{0.5cm}
\begin{minipage}[c]{7.5cm}
\epsfig{file=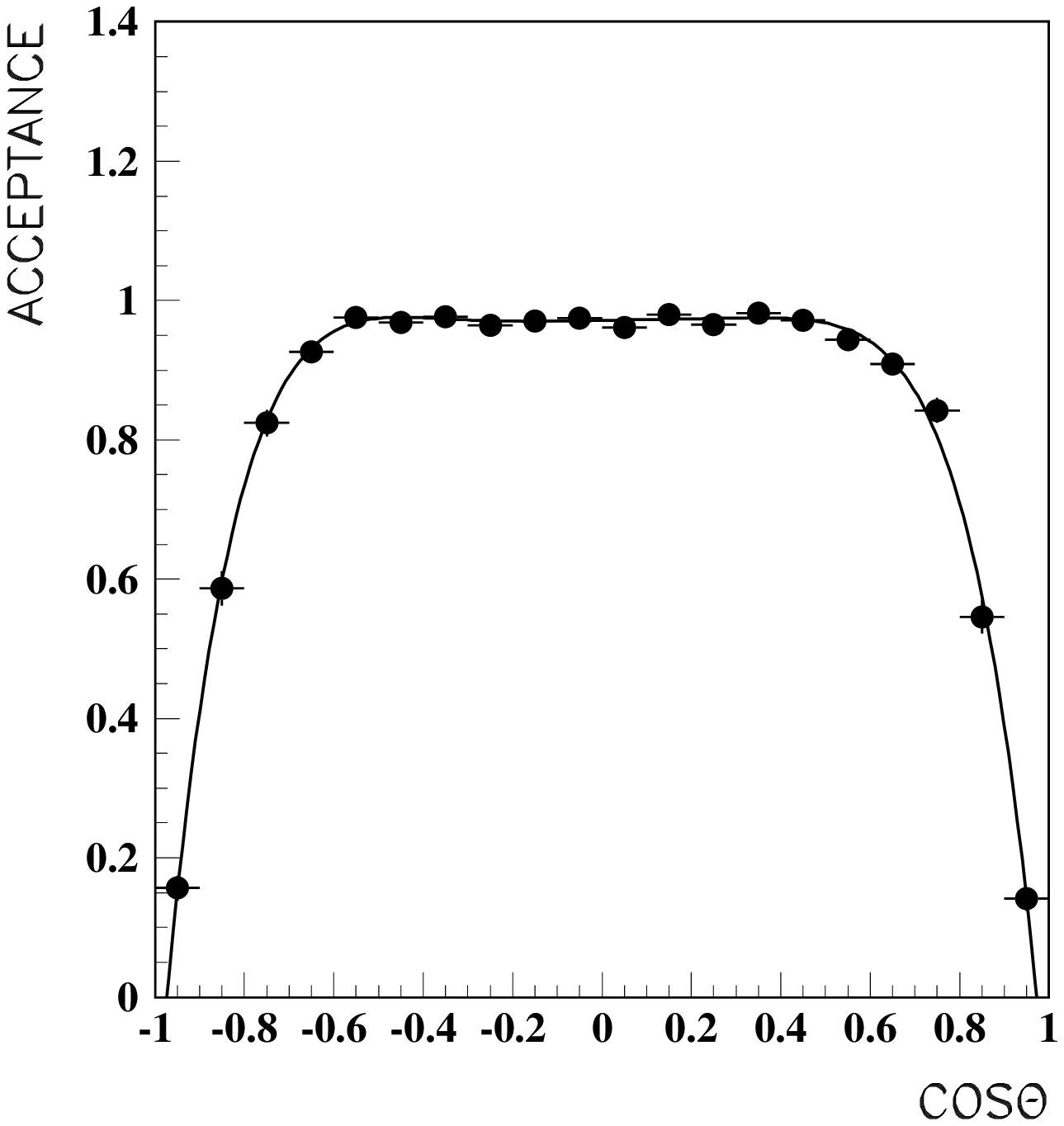,height=7.5cm,width=7.5cm}
\caption{Acceptance for the $\Upsilon$ events
as a function of the CMS polar angle}
\label{accept}
\end{minipage}
\end{figure}

In the following only the muonic decay channel is considered because the
expected mass resolution is better.
Figure 1 shows the $\Upsilon \to \mu^+\mu^-$ mass distributions
on top of the Drell-Yan pair production expected
in one year $(10^7 s)$ of HERA-$B$ running with $40$ MHz interactions
rate.  
Cut on the muon momenta,  $p>5$ GeV/c and $p_T>0.5$ GeV/c, are applied to 
satisfy the trigger requirements. The expected statistics are about 10000
fully reconstructed $\Upsilon$ events per year.
The background is largely dominated by the Drell-Yan process.
It is simulated by PYTHIA with the cross section calculated
at the leading order, thus a $K$ factor of $2.3$ is taken into account
according to \cite{exp2}.
As can be seen from Fig.1 the predicted  mass 
resolution of about $1\%$ and a signal to background ratio S/B$\simeq1.3$
(in a $\pm2\sigma$ mass window around the $\Upsilon(1S)$)
allows to clearly observe the $\Upsilon$
mass peaks above the background and to separate the $\Upsilon(1S)$ and
$\Upsilon(2S)$ states.

Figure 2 shows the acceptance for $\Upsilon$ events as a function
of the cosine of the polar angle between the positive muon momentum 
and the beam direction in the rest frame of $\Upsilon$.
In almost the full range of $\cos\theta$ the acceptance is close to unity.
Large acceptance corrections are expected only for $|\cos\theta|>0.8$.

The accuracy of the polarization measurement is estimated by
simulating various initial polarizations for $\Upsilon$ mesons and
Drell-Yan pairs with various signal to background ratios.
After corrections for the acceptance
the resulting angular distributions are fitted with the function:
\begin{equation}
\frac{dN}{d\cos\theta}=S(1+\alpha\cos^2\theta)+B(1+\alpha^\prime\cos^2\theta).
\end{equation}
Here $S$ and $B$ are signal and background normalization factors
and $\alpha$ and $\alpha'$ determine corresponding polarization parameters.
This procedure allows to extract the polarization not only of $\Upsilon$
mesons but also of the Drell-Yan pair, which is an interesting issue by itself.
Assuming that
Drell-Yan polarization and signal to background ratio could be precisely
measured in the HERA-$B$ experiment, the number of free parameters in the
fitting function is reduced to 2. Results on the parameter $\alpha$ 
obtained from the fit for the different MC input values and for 10000
reconstructed $\Upsilon$ events are shown in table III.

\begin{center}
\begin{tabular}{|l|c|c|c|c|}
\hline
Input values& \multicolumn{2}{|c|}{S/B=1.3} & S/B=1 & S/B=2\\
\cline{2-5} 
\hspace{0.5cm} of $\alpha$ & $\alpha'=0$ & $\alpha'=1$ & 
\multicolumn{2}{|c|}{$\alpha'=1$}\\
\hline
\hline   
$\alpha=0.25$ & $0.21\pm0.08$ & $0.21\pm0.08$ &$0.22\pm0.09$ & $0.23\pm0.08$ \\
\hline
$\alpha=0.5$ & $0.52\pm0.10$ & $0.48\pm0.09$ &$0.51\pm0.09$ & $0.54\pm0.09$ \\
\hline
$\alpha=1$ & $0.91\pm0.12$ & $0.96\pm0.11$ &$0.96\pm0.12$ & $0.98\pm0.10$ \\
\hline
\end{tabular}
\end{center}

\noindent
{\bf Table III.} Parameter $\alpha$ obtained from the fit for different
MC input values.
\vspace{0.3cm}

We note that the expected statistical error on $\alpha$ is largely 
dominated by the $\Upsilon$ statistics rather than by the value of 
the background
polarization and the signal to background ratio.
Figure 3 illustrates the statistical error $\delta\alpha$ as a function 
of the number of reconstructed $\Upsilon$ events 
\begin{wrapfigure}{l}{7.5cm}
\vspace{0.5mm}
\centering
\epsfig{file=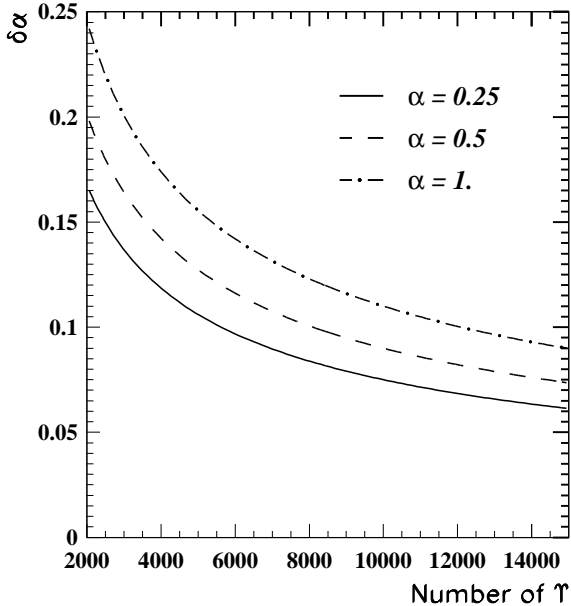, width=7.5cm}
\vspace{-7mm}
\caption{\small 
Attainable accuracy in $\alpha$ for $\alpha=$0.25, 0.5 and 1
vs. the number of the reconstructed $\Upsilon$ events. S/B=1.3}
\vspace{-1.5cm}
\end{wrapfigure}
for various $\alpha$'s.
As can be seen, an accuracy on the polarization parameter
$\delta\alpha\simeq0.08$ for $\alpha=0.25$ can be achieved for one year
of the HERA-$B$ running. The expected error will allow to distinguish
between
$\alpha=0$ (CEM) and $\alpha\simeq0.24-0.3$ (NRQCD FA)
with a $3\sigma$ significance.

\section{Conclusions}

The polarization of $\Upsilon$ mesons is calculated at fixed target
energies ($\sqrt{s}\simeq40$ GeV). It is shown that in the NRQCD
factorization approach
$\Upsilon$ mesons are expected to be produced transversely polarized; the 
parameter $\alpha$ for the polar angle distribution of quarkonium decay
products is about $0.24\div0.3$. In contrast, the color evaporation model 
postulates that multiple soft gluon exchange in the hadronization phase
destroys the initial polarization of the heavy quark pair and quarkonium
is produced unpolarized.

Higher twist effects are expected to be small due to the 
large mass of the $b$-quark. The contribution of higher Fock states 
in bottomonium production are more suppressed than in the charmonium
case, the relative velocity for the bottomonium family is about $v^2\simeq0.1$.
Thus the measurement of the $\Upsilon$ polarization provides an excellent 
opportunity to test different mechanisms of heavy quarkonium production. 
In particular, it allows to distinguish between the NRQCD FA \cite{NRQCD}
and the CEM \cite{CEM}.
On the other side, the observation of an extremely large polarization will
indicate
that $\Upsilon$ mesons are mainly produced through color singlet states and 
that the color octet parameters for $\chi_{bJ}$  production extracted from
the Tevatron data should be much smaller than known at present. 

The Monte Carlo simulation shows that the projected statistical error for
the measurement of
the polarization parameter  $\alpha$ is about $0.08$ for $\alpha\simeq0.25$ 
in one year running of the HERA-$B$ experiment. The simulation
is done only for the $\mu^+\mu^-$ decay channel. A statistics gain of
almost
a factor 2 is expected from the $e^+e^-$ channel. The simulation 
conservatively assume the lowest value for the $\Upsilon$ production 
cross section measured 
in different experiments \cite{exp1,exp2,exp3} at $\sqrt{s}\simeq38.8$ GeV
which corresponds to a HERA proton beam momentum of $800$ GeV/c 
(the current value for HERA is 920 GeV/c).       

\bigskip
 
{\Large \bf Acknowledgments}
 
\medskip
We are grateful to S.~Brodsky, R.~Mankel and M.~V\"anttinen  for useful 
comments and discussions.
We thank  W.-D.~Nowak for helpful comments and careful reading of 
this manuscript. 
A.T. acknowledges  the  support  by the Alexander von
Humboldt Foundation.  A.S. is grateful to the DFG financial
support (III GK - GRK 271/1).

\end{document}